\documentclass[11pt,twoside]{article}
\usepackage{asp2010}
\usepackage{graphicx}

\resetcounters

\begin{document}

\title{Variability in Quasar Broad Absorption Line Outflows}
\author{Daniel M. Capellupo,$^1$ Fred Hamann,$^1$ Joseph C. Shields,$^2$ Jules Halpern,$^3$ Paola Rodr\'iguez Hidalgo,$^4$ and Tom A. Barlow$^5$
\affil{$^1$Department of Astronomy, University of Florida, 211 Bryant Space Science Center, Gainesville, FL 32611-2055}
\affil{$^2$Department of Physics \& Astronomy, Ohio University, Clippinger Labs 251B, Athens, OH 45701}
\affil{$^3$Department of Astronomy, Columbia University, 550 West 120th Street, New York, NY 10027}
\affil{$^4$Department of Astronomy and Astrophysics, Pennsylvania State University, 525 Davey Lab, University Park, PA 16802, USA}
\affil{$^5$Infrared Processing and Analysis Center, California Institute of Technology, 1200 E. California Blvd., Pasadena, CA 91125, USA}}

\begin{abstract}
Broad absorption lines (BALs) in quasar spectra identify high velocity outflows that likely exist in all quasars and could play a major role in feedback to galaxy evolution. Studying the variability in these BALs can help us understand the structure, evolution, and basic physical properties of these outflows. We are conducting a BAL monitoring program, which so far includes 163 spectra of 24 luminous quasars, covering time-scales from $\sim$1 week to 8 years in the quasar rest-frame. We investigate changes in both the CIV $\lambda$1550 and SiIV $\lambda$1400 BALs, and we report here on some of the results from this program.
\end{abstract}

\section{Introduction}

High-velocity wind-like outflows are an important part of the quasar system and a potential contributor to feedback to the host galaxy. Many properties of quasar outflows are still poorly understood, including their location and three-dimensional structure. One way to obtain constraints on quasar outflows is to study the variability in their absorption lines. Variability on shorter time-scales can place constraints on the distance of the absorbing material from the central supermassive black hole (SMBH). Measurements of variability on longer (multi-year) time-scales provide insight into the homogeneity and stability of the outflowing gas. Overall, results of variability studies provide information on the size, kinematics, and internal makeup of sub-structures within the outflows.

Broad absorption lines (BALs) are the most prominent signatures of accretion disk outflows seen in quasar spectra. We are conducting a BAL monitoring program for a sample of 24 luminous BAL quasars at z=1.2$-$2.9 from \citet{Barlow93}. We have been re-observing these quasars at the MDM Observatory 2.4-m Hiltner and KPNO 2.1-m telescopes, and we also include spectra from SDSS when available. We currently have in total 163 spectra of these 24 quasars.

To identify BAL variability, we first define the regions of BAL absorption, based on the definition of the `balnicity index' (BI), i.e. they must contain contiguous absorption that reaches $\geq$10 per cent below the continuum across $\geq$2000 km/s \citep{Weymann91}. We then used visual inspection to identify velocity intervals with a width of at least 1200 km/s that varied. Using the average flux and associated error within each candidate variability interval, we place an error on the flux differences between the two spectra. We include all intervals of variability where the flux differences are at least 4$\sigma$.

\section{CIV and SiIV BAL Variability}

\subsection{Trends in the CIV Short-Term and Long-Term Data}

\begin{figure}
 \begin{center}
  \includegraphics[scale=0.48]{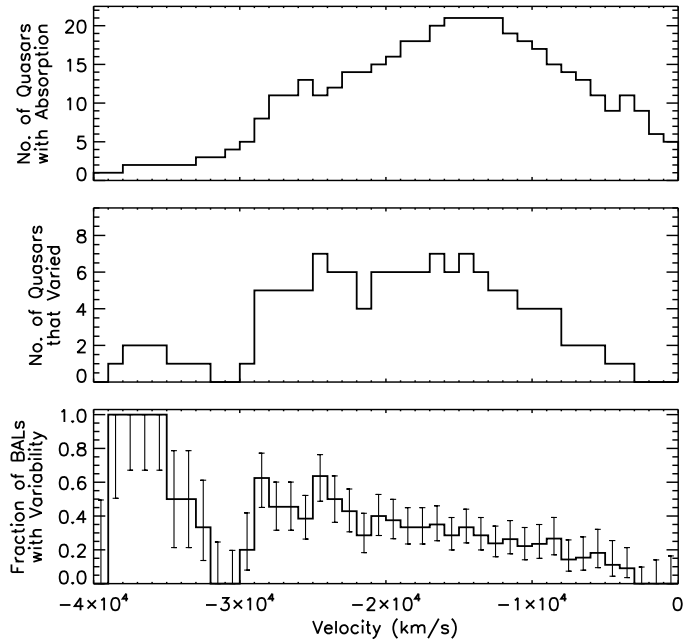}
 \end{center}
  \caption{The top two panels show the number of occurrences of CIV BAL absorption and BAL
    variability versus velocity. The third panel is the second panel divided by the first, with 1$\sigma$
    error bars overplotted.}
  \label{hist}
\end{figure}

The first results of our BAL variability study were reported in \citeauthor{Capellupo11a} (\citeyear{Capellupo11a}; hereafter, Paper 1), where we focused on variability in just the CIV $\lambda$1550 BALs in two different time intervals: 4$-$9 months (short-term) and 3.8$-$7.7 yr (long-term) in the quasar rest-frame. We found both a higher incidence of variability (65\%, or 15/23, of the quasars versus 39\%, or 7/18) and a larger typical change in strength in the long-term data. The variability occurs typically in only portions of the BAL troughs (e.g. Fig. \ref{sp1246}; see also, \citealt{Gibson10}). In Fig. \ref{hist}, we plot the incidence of CIV BAL absorption and variability versus velocity. The BAL components at higher outflow velocities are more likely to vary than those at lower velocities. We also find in Paper 1 that weaker BALs are more likely to vary than stronger BALs.

\subsection{Multi-Epoch Monitoring of SiIV and CIV Variability}

In \citeauthor{Capellupo11b} (\citeyear{Capellupo11b}; hereafter, Paper 2), we directly compare the variability of SiIV $\lambda$1400 BALs to CIV in the long-term dataset from Paper 1. C and Si have different abundances, if solar abundances are assumed, and they have different ionization properties (e.g. \citealt{Hamann08}). By examining if CIV and SiIV have different variability properties, and how they differ, coupled with these differences in abundances and ionization properties, we can gain new insight into the cause(s) of BAL variability.

SiIV BALs are more likely to vary than CIV BALs. For example, when looking at flow speeds $>$$-$20 000 km/s, 47 per cent of the quasars in our sample exhibited SiIV variability while 31 per cent exhibited CIV variability. Variability in SiIV can occur without corresponding changes in CIV at the same velocities. For $\sim$50 per cent of the variable SiIV regions, there was no corresponding CIV variability at the same velocities. However, we have only one tentative case where changes in CIV are not matched by SiIV. At BAL velocities where both CIV and SiIV varied, the changes always occurred in the same sense (with both lines either getting stronger or weaker).

We now have up to 13 epochs of data per object for our sample of 24 BAL quasars. With all the observing epochs included, the fraction of quasars with CIV BAL variability at any velocity increases from the value of 65\% found in Paper 1 to 88\% for the same sample of quasars. This increase was caused by variations missed in the 2-epoch comparisons in Paper 1. We also find that BAL changes at different velocities in the same ion {\it almost} always occurred in the same sense. We found just 3 cases that show evidence for one CIV BAL weakening while another strengthens within the same object. The multi-epoch data also show that the BAL changes across $\sim$1 week to 8 years in the rest-frame were not generally monotonic (see also, \citealt{Gibson10}). Thus, the characteristic time-scale for significant line variations, and (perhaps) for structural changes in the outflows, is less than a few years. Furthermore, with more epochs added, we still do not find clear evidence for acceleration or deceleration in the BAL outflows.

\subsection{Time-Scales of Variability}

We take a closer look at the time-scales of variability in \citeauthor{Capellupo12} (\citeyear{Capellupo12}; hereafter, Paper 3). In Spring 2010, we re-observed a subsample of our BAL quasars multiple times over rest-frame time intervals of $\Delta$t $\sim$1 week to 1 month to augment our temporal sampling at these short time-scales. Variability results on the shortest time-scales are important for putting constraints on the location and sizes of the outflows.

\begin{figure}[ht]
 \begin{center}
  \includegraphics[scale=0.45]{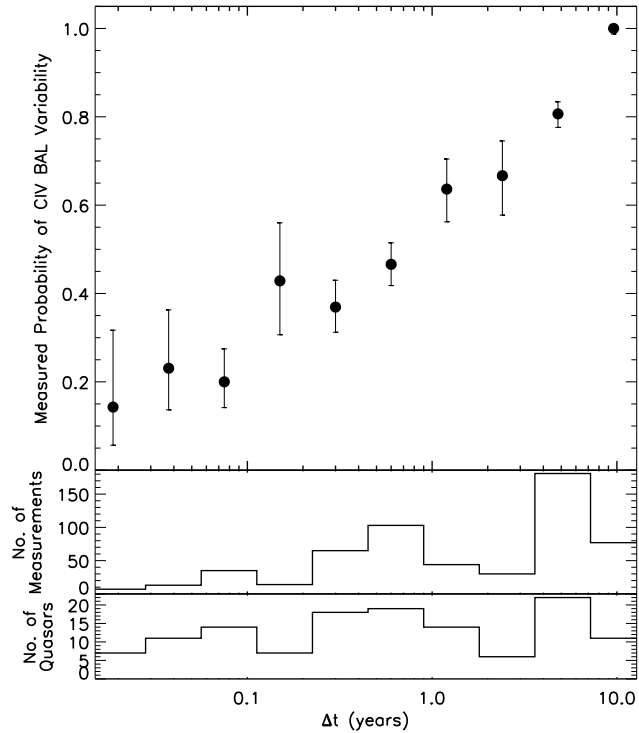}
 \end{center}
  \caption{The top panel displays the fraction of measurements where there was CIV BAL variability
    versus the time interval between the two observations. The middle and bottom panels show the
    number of measurements and the number of quasars, respectively, that contribute to each bin.}
  \label{dream}
\end{figure}

The multi-epoch nature of our dataset, spanning time intervals from 0.02 to 8.7 yr in the rest frame, allows us to investigate the probability of BAL variability versus $\Delta$t. We examined CIV BAL absorption at all measured velocities in our entire data sample. We compared each pair of observations for each quasar and counted the occurrences of CIV BAL variability, using our definition of BAL variability defined in Paper 1 (see Section 1 above). We then calculated a probability by dividing the number of occurrences of variability by the number of measurements, where a pair of observations is one measurement, in logarithmic bins of $\Delta$t. We plot this measured probability of detecting CIV BAL variability versus $\Delta$t in Fig. \ref{dream}. The middle and bottom panels display the number of measurements and the number of quasars, respectively, that contribute to each $\Delta$t bin. Each quasar can contribute multiple times to each bin.

\begin{figure}
 \begin{center}
  \includegraphics[scale=0.46, angle=90]{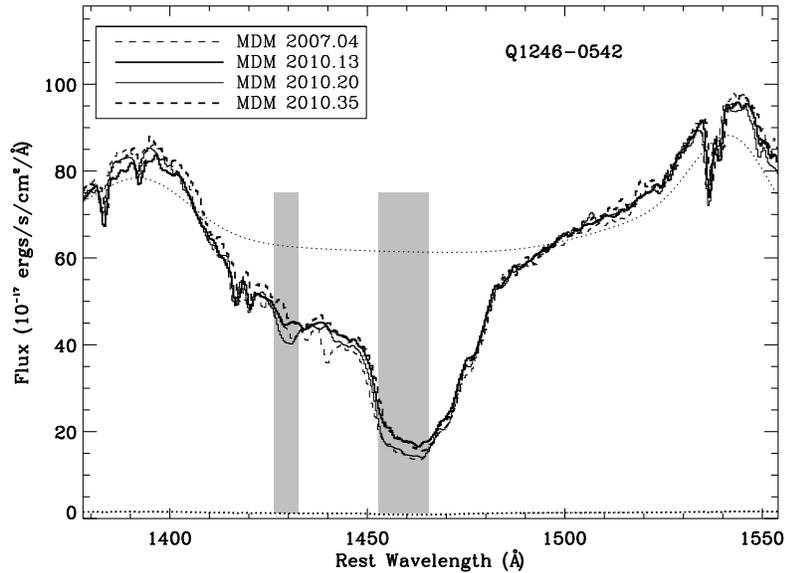}
 \end{center}
  \caption{Smoothed MDM spectra of 1246$-$0542, showing the CIV absorption region. The bold
    and thin solid curves show the two epochs separated by 8 days in the rest-frame, and the shaded
    regions indicate the velocity intervals that varied between these two epochs. The formal
    1$\sigma$ errors are shown near the bottom.}
  \label{sp1246}
\end{figure}

In our dataset, we found variability down to a $\Delta$t of 8 days in 1246$-$0542 between the 2010.13 and 2010.20 observations. Fig. \ref{sp1246} shows the CIV region of the spectra for these two observations as bold and thin lines, respectively. As indicated by shaded bars in Fig. \ref{sp1246}, there are two separate velocity intervals of variability between the two spectra, and the flux difference is at least 8$\sigma$ in each interval. We also overplot an earlier (2007.04; thin dashed curve) and a later (2010.35; bold dashed curve) epoch in Fig. \ref{sp1246}, which shows that there was variability in these intervals in other epochs as well.

\section{Conclusions}

Coordinated variabilities between absorption regions at different velocities in individual quasars seems to favor changing ionization of the outflowing gas as the cause of the observed BAL variability. Many of our results are consistent with this scenario, including our findings that SiIV, which has a lower optical depth and is therefore less likely to be saturated than CIV, is more likely to vary than CIV. However, variability in limited portions of broad troughs fits naturally in a scenario where movements of individual clouds, or substructures in the flow, across our lines-of-sight cause the absorption to vary. This scenario is also consistent with the main results of Paper 2 (Section 2.2 above) if SiIV has a smaller covering fraction than CIV. The actual situation may be a complex mixture of changing ionization and cloud movements (Paper 2).

If the short time-scale variations in 1246$-$0542 were due to cloud movements and a $\sim$10\% change in line strength over 8 days indicates the cloud crossed 10\% of the continuum source in that time, then the transverse speed is $\sim$$-$27 000 km/s. If we assume the transverse speed is related to the Keplarian rotation speed, then the outflowing gas is located $\ll$1 pc from the central SMBH, probably inside the radius of the broad emission-line region (Paper 3).

\acknowledgements This work was supported by grant number 1009628 from the National Science Foundation, and by the Guest Observer program 11705 with the Space Telescope Science Institute.

\bibliography{bibliography}

\end{document}